\documentclass[10pt]{iopart}
\usepackage{graphicx}
\usepackage{latexsym}
\usepackage{ulem}
\usepackage{dcolumn}
\usepackage{bm}
\usepackage{color}
\usepackage{amsfonts}

\newcommand{\be}{\begin{equation}}
\newcommand{\ee}{\end{equation}}
\newcommand{\bea}{\begin{eqnarray}}
\newcommand{\eea}{\end{eqnarray}}
\newcommand{\bse}{\begin{subequations}}
\newcommand{\ese}{\end{subequations}}

\renewcommand{\comment}[1]{}

\begin{document}

\paper[]{Coupled transport in a linear-stochastic Schr\"odinger Equation}
\date{\today}

\author{Stefano Iubini$^{1,2}$}

\address{$^{1}$ Dipartimento  di Fisica e Astronomia, Universit\`a di Padova, Via Marzolo 8, I-35131 Padova, Italy}
\address{$^{2}$ Istituto dei Sistemi Complessi, Consiglio Nazionale
delle Ricerche, via Madonna del Piano 10, I-50019 Sesto Fiorentino, Italy
}

\ead{stefano.iubini@unipd.it}

\begin{abstract}
I study heat and norm transport in a one-dimensional lattice of linear Schr\"odinger oscillators with conservative stochastic perturbations.
Its equilibrium properties are the same of the Discrete Nonlinear Schr\"odinger
equation in the limit of vanishing nonlinearity.
When attached to external classical reservoirs that impose nonequilibrium conditions, the chain  displays diffusive transport, with finite Onsager coefficients in the thermodynamic limit and a finite Seebeck coefficient.

\end{abstract}

\submitto{Journal of Statistical Mechanics: theory and experiment}

\section{Introduction}

Within the vast class of nonequilibrium classical and quantum phenomena, the physics of coupled transport (CT)
is a growing field with potentially revolutionary technological innovations.
The knowledge of systems where two or more species of irreversible flows may occur and influence
 one another dates back to the discoveries by Seebeck and Peltier of thermoelectricity in
the first half of the XIX century. The basic principle of a thermoelectric material is the capacity of
converting a temperature difference into a voltage and vice-versa. More generally, depending on the
physical nature of the microscopic carriers, analogous CT effects may also involve different kind of
unbalance, like chemical potential differences (thermodiffusion) or spin voltages (thermomagnonics).

From the point of view of applications, the possibility to control a heat flow by means of an auxiliary
current within the same medium allows  to build solid-state miniaturized versions of power
generators or refrigerators with no mechanical parts. At present, however, the conversion efficiency of these materials is
still too low compared to the one of common thermo-mechanical machines. As a result, the use of CT devices
is limited to a quite restricted number of special applications.

In the recent years, mounting evidence has been provided that the aim of a significant improvement of CT
technologies necessarily requires a deeper understanding of the fundamental physical
mechanisms that underlie CT processes~\cite{benenti2017fundamental,luo2018thermodynamic,benenti2013conservation,mejia2001coupled,larralde2003transport}. In the absence of a complete theory, many approaches have been
proposed to tackle the problem. In the context of statistical mechanics, the study of simple models of coupled oscillators
represents undoubtedly a powerful strategy for unveiling the basic microscopic mechanisms of transport phenomena~\cite{Lepri2016,LLP03,DHARREV}.
Among them, the Discrete Nonlinear Schr\"odinger (DNLS) equation~\cite{Kevrekidis}  is a natural candidate to study coupled transport of energy and mass (norm) in
several setups, ranging from nonlinear optics~\cite{jensen1982,christodoulides1988} to cold atoms~\cite{trombettoni2001,livi2006,hennig2010transfer} and micromagnetic systems~\cite{borlenghi14,borlenghi15}.  In one dimension, the DNLS equation reads
\begin{equation}
i \dot {z}_n = -\nu |z_n|^2z_n - z_{n+1}-z_{n-1} 
\label{eq:dnls}
\end{equation}
It describes the dynamics of a chain of $N$ coupled anharmonic  oscillators (with $1\leq n \leq N$) with complex amplitudes $z_n$ ($|z_n|^2$ is the local norm)  and a real nonlinearity coefficient $\nu$.
In the standard nonequilibrium setup, the DNLS chain interacts with two external reservoirs that exchange energy and norm and impose
temperature and chemical potentials~\cite{iubini2012,iubini2013}. 
While for sufficiently large temperatures, the DNLS model displays normal transport with a non-vanishing Seebeck coefficient~\cite{iubini2012,Iubini2016} and diffusive spreading of energy- and norm correlations~\cite{Mendl2015}, in the low-temperature regime,
the quasi-conservation of the phase differences yields anomalous transport on long time scales~\cite{Mendl2015}.
More in general, it was shown that the nonlinearity
of the system plays a relevant role for the determination of its nonequilibrium properties. Some examples are the strong dependence of the
Onsager coefficients on the thermodynamical variables~\cite{iubini2012}, the observation of interfacial regions and  dynamical transitions
for low temperatures and large chemical potential differences~\cite{iubini2014} and the spontaneous creation of nonequilibrium  barriers and localized structures for very large (even negative) temperatures~\cite{wall17}. 

For vanishing nonlinearity, the DNLS equation reduces to a chain of complex harmonic oscillators, the Discrete Schr\"odinger  (DS) equation. In this limit transport is 
ballistic (i.e. currents do not depend on the system size) and stationary profiles are flat~\cite{iubini2012}, in analogy with the well known behaviour of a chain of real
harmonic oscillators connected to boundary reservoirs~\cite{RLL67}. Since in the DS equation energy and norm are carried by $N$ 
noninteracting phonon modes, the related currents are independent on the system size $N$ and can be fully characterized
in the framework of Landauer theory of electronic transport~\cite{sheng2006,Dhar2006}.

A particularly effective strategy to bridge the gap between nonlinear  models and  harmonic systems is to replace nonlinear interactions by suitable stochastic   perturbations of a linear dynamics~\cite{BBO06}.
 This approach proved to be very effective in the context of anomalous heat conduction,
allowing to obtain explicit representations of the nonequilibrium invariant measure~\cite{DLLP08} and
 analytical expressions  of stationary temperature profiles associated to anomalous currents~\cite{Lepri2009,Lepri2010,Delfini10}.	
In a few words, the method consists in adding local  stochastic collisions that conserve exactly momentum and energy of a chain 
of harmonic oscillators.
Such collisions mimic the effect of nonlinearities and introduce some degree of ergodicity and irreversibility
that would otherwise be completely missing in a purely harmonic system.   
It is important to notice that as long as one considers a chain of real harmonic oscillators, the nature of the interaction potential and the symmetries of the model prevent any coupling between momentum- and heat currents~\cite{Iubini2016,spohn2014fluctuating}. As a result, no coupled transport is expected in this context.

Analogous studies focused on coupled transport are scarce~\cite{olla2019fourier}.
For the DS chain, it was shown in~\cite{letizia2017diffusive} that the addition of phase noise to the deterministic dynamics 
allows to recover nonequilibrium stationary states that satisfy the Fourier law. This class of perturbations, however, 
conserves only the total norm, while the conservation of the total energy is lost. As a result, the exploration of the transport 
properties of the system is limited to the set of states at infinite temperature (and infinite chemical potential)~\cite{letizia2017diffusive,iubini2013}.

In this paper I present a discrete linear-stochastic 
Schr\"odinger equation that conserves exactly both the total energy and the total norm and naturally  allows to study coupled transport problems in the whole space of thermodynamic variables. 
 This model has the 
same symmetries of the DNLS equation  and
recovers its typical transport properties, which are characterized by diffusive currents and a nonzero Seebeck coefficient. 
The paper is organized as
follows. In section 2, I introduce the model and discuss the relevant thermodynamic observables that are used to characterize nonequilibrium steady states. In section 3, I discuss the main features of the nonequilibrium steady states as obtained from 
numerical simulations. Finally, section 4 is devoted to the conclusions and to a discussion of open problems and future perspectives.

\section{Model and observables}

I begin this section by reviewing the equilibrium properties of the one dimensional  DS equation
\begin{equation}
i \dot {z}_n =  - z_{n+1}-z_{n-1} \; 
\label{eq:ds}
\end{equation}
Upon identifying the set of canonical variables $z_n$ and $i z_n^*$, equation (\ref{eq:ds}) can be derived from 
the Hamilton equations $\dot{z}_n = - \partial H / \partial (i z_n^*)$ for the Hamiltonian
\begin{equation}
 H= \sum_{n=1}^{N} \left( z_n^*z_{n+1}+z_nz_{n+1}^* \right) 
 \label {Hz}
 \end{equation}
The model has two exactly conserved quantities, namely the total energy $H$ and the total norm 
\begin{equation}
A = \sum_{n=1}^N |z_n|^2
\end{equation}
In terms of Fourier amplitudes $\tilde{z}_k=\frac{1}{\sqrt{N}} \sum_{n=1}^N e^{-\frac{2\pi i}{N}kn} z_n$ (with $ k=1,\ldots N$),
the Hamiltonian can be rewritten in the diagonal form
\begin{equation}
 H=\sum_{k=1}^N |\tilde{z}_k|^2 \omega_k \quad,
\end{equation}
where $\omega_k=2\cos(2\pi k/N)$ is the energy spectrum of the system and $|\tilde{z}_k|^2$ represents the
power of the $k$-th mode, with the constraint $\sum_{k=1}^N |\tilde{z}_k|^2=A$. 

\begin{figure}
\begin{center}
\includegraphics*[width=0.8\textwidth]{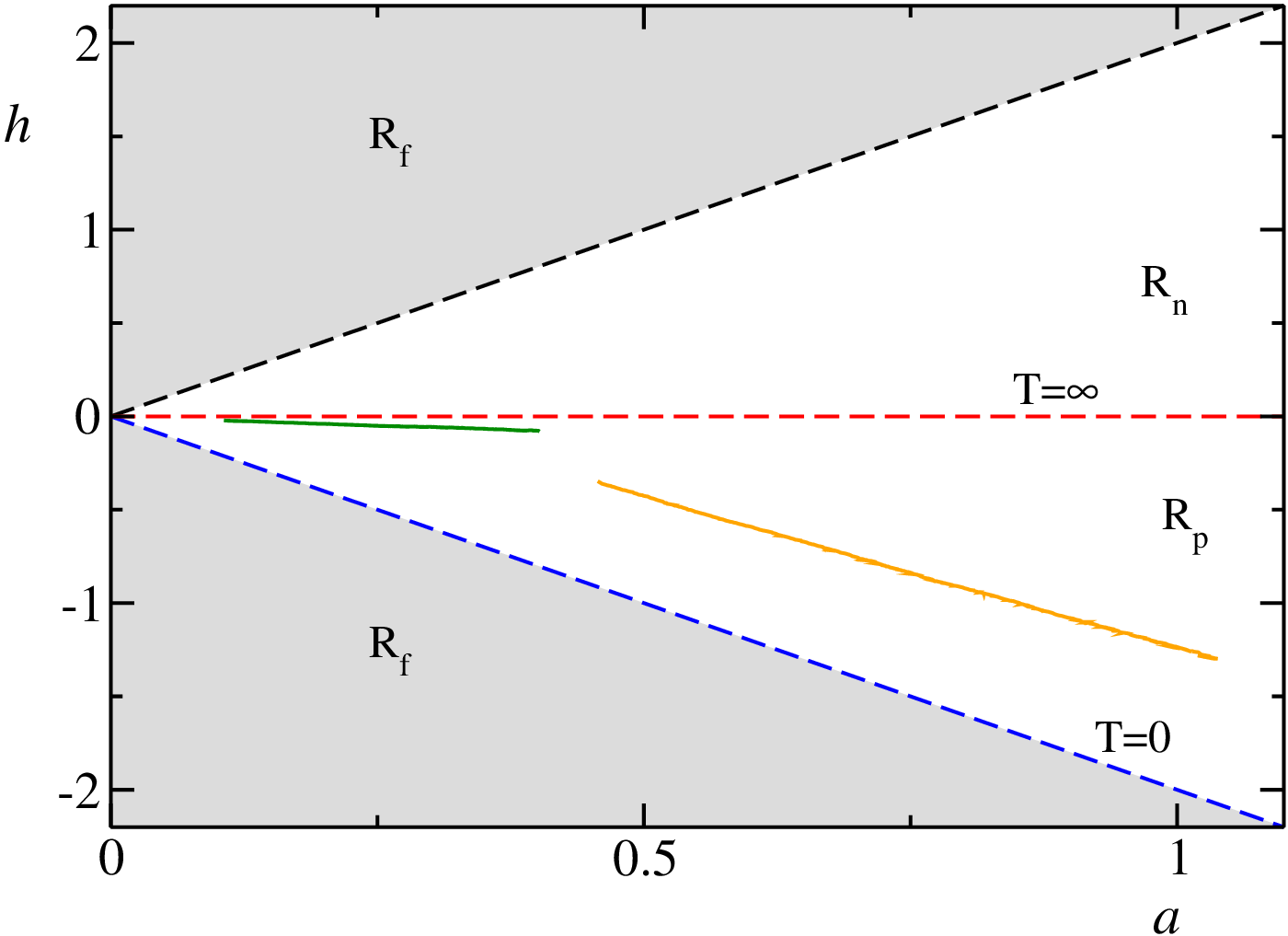}
\end{center}
\caption{Equilibrium phase diagram (a,h) of the DLS equation. The blue  dashed curve refers to the $T=0$ isothermal while
the red dashed line identifies the isothermal $T=\infty$.  Solid lines show the nonequilibrium profiles of local
norm and local energy $[a(x),h(x)]$ in the presence of conservative noise and  $T_L=1$, $T_R=4$, $\mu_L=\mu_R=-10$ (green line)
and $T_L=T_R=1$, $\mu_L=-2.2$, $\mu_R=-3$ (orange line).
}
\label{f:fig1}
\end{figure}

The thermodynamical properties of the DS equation are determined by two parameters: the total norm density $a=A/N$ 
and the total energy density $h=H/N$, which can be mapped to a couple of values of temperature $T$ and chemical 
potential $\mu$ via a classical grand-canonical distribution. To derive this mapping, one can start from the partition
function of the system 
\begin{equation}
\mathcal{Z}=\int \prod_{k=1}^N d\tilde{z}_k\,d\tilde{z}_k^* \,e^{-(H-\mu A)/T}\quad,
\label{eq:Z1}
\end{equation}
which can be explicitly computed and rewrites in the standard equipartition form
\begin{equation}
 \mathcal{Z}=\prod_{k=1}^N \frac{2\pi}{\beta(\omega_k-\mu)}\quad\
 \label{eq:Z2}
\end{equation}
where  the Boltzmann constant was set to 1. 
Hence, from the knowledge of the free energy  $\mathcal{F}=-T \log(\mathcal{Z})$, one can derive~\cite{Rumpf2004} an equation of state for $a$ and $h$ in the form
\begin{eqnarray}
a&=&-\frac{T \; \mathrm{sign}(\mu)}{\sqrt{\mu^2-4}} \nonumber\\
 h_&=& T +\mu a \label{eq:s2}
\end{eqnarray}
with the condition $|\mu|>2$\footnote{This condition ensures the existence of a well defined partition function in (\ref{eq:Z1}) and (\ref{eq:Z2})}. Equations  (\ref{eq:s2}) fully specify the equilibrium 
phase diagram of the model, as shown in figure~\ref{f:fig1}. 
Here, the lower dashed line represents the ground state $h=-2a$, while the horizontal dashed line $h=0$ identifies the set of 
states at infinite temperature. The grey region $R_f$  is a forbidden region, while $R_p$ is
the region of positive  temperature states. In analogy to what was found for the DNLS equation~\cite{Rasmussen2000}, the DS equation displays a region $R_n$ of
negative temperature states located above the infinite temperature line. The symmetric nature of $R_n$ and $R_p$ (which is found also
in some spin systems~\cite{ramsey1956}) derives from the fact
that for every accessible state $(a,h)$ corresponding to some temperature $T$, there exists a symmetric state $(a,-h)$ characterized by the temperature $-T$. In view of this symmetry,  in this paper I will focus to the nonequilibrium 
properties of the DS model within the region $R_p$. A discussion of more general situations involving both $R_p$ and $R_n$ will be given in a forthcoming paper.

\subsection{Conservative noise}

The stochastic version of the DS equation is obtained from the deterministic one, equation  (\ref{eq:ds}), by adding suitable 
conservative random ``shakings'' of local phases, which occur at rate $\gamma_c$. 
To illustrate this stochastic dynamics, let me consider the local energy on a generic site $n$
 \begin{equation}
 e_n = 2\rho_n \rho_{n-1} \cos(\phi_n-\phi_{n-1}) + 2\rho_n\rho_{n+1} \cos(\phi_{n+1}-\phi_n) 
 \label{eq:hn}
 \end{equation} 
 where $\rho_n=|z_n|$ and $\phi_n=\arg(z_n)$. The goal is to define a local transformation on the lattice site $n$ that
 conserves both the energy $e_n$ and the local norm $\rho_n^2$. If the transformation restricts to the phase $\phi_n$, 
 the conservation of the  norm holds straightforwardly, while  from equation (\ref{eq:hn}) it follows the condition
 \begin{equation}
  A\cos(\phi_n - \alpha) +B\cos(\phi_n-\beta) = E
 \end{equation}
 where $A=\rho_{n-1}$, $B=\rho_{n+1}$, $\alpha=\phi_{n-1}$, $\beta=\phi_{n+1}$ and $E$ are now fixed parameters. This linear
 goniometric equation has two (possibly identical) solutions in the interval $[0,2\pi)$: one solution corresponds to the state before the shaking operation, the other one to the shaken state. Shaking operations are performed on the whole DS  chain at random times, whose 
 separation $\tau$ are independent and identically distributed variables extracted from a Poissonian distribution $P(\tau)\sim \exp(-\gamma_c \tau)$.  In between two successive realizations of the conservative noise, the evolution of the system follows the deterministic dynamics defined in 
 equation~(\ref{eq:ds}), except for the two boundary sites, which are coupled to external stochastic reservoirs. Their 
dynamics is described in 
 the following subsection, along with a presentation of the nonequilibrium observables that are considered in this paper.
 
\subsection{The nonequilibrium setup}  

 A typical setup adopted for the study of coupled transport problems amounts to
   put the system in contact with two boundary reservoirs
 at temperature $T_L$ and $T_R$ and chemical potential $\mu_L$ and $\mu_R$, respectively. For the stochastic DS equation this task can be accomplished by introducing 
 a suitable Langevin equation~\cite{iubini2013}. Focusing on the last lattice site $n=N$, which is in contact with the right reservoir, 
 the Langevin equation writes
 \begin{equation}
  i \dot z_{N}= -(1+i \gamma) z_{N-1}  
 +i\gamma \mu_R z_{N}+ \sqrt{\gamma T_R} \, \eta(t) \quad  \; ,
 \end{equation}
where $\eta(t)$ is a complex Gaussian white noise with zero mean and unit variance, $\gamma$ is the bath coupling parameter and 
open boundary conditions are assumed. An analogous equation holds for the left reservoir, which is coupled to the first site $n=1$ and
imposes a temperature $T_L$ and a chemical potential $\mu_L$.

When the chain is steadily kept out of equilibrium by some thermodynamic force, the main  observables are the two currents associated to the conserved quantities: the norm flux $j_a$ and the energy flux $j_h$. Their explicit expressions are 
\begin{eqnarray}
j_a &=& 2 \langle Im(z_n^* z_{n+1}) \rangle \label{eq:ja}  \\
j_h &=& 2 \langle Re(\dot{z}_n z_{n+1}^*) \label{eq:jh}\rangle 
\end{eqnarray}
where the angular brackets denote a time average.  Within the linear response regime, thermodynamic forces and currents are
related by the celebrated Onsager relations~\cite{Saito2010,iubini2012,Iubini2016}. 
Upon introducing the heat flux $j_q=j_h-\mu j_a$, they write
\begin{eqnarray}
j_a &=& -L_{aa} \beta\frac{d \mu}{dx} + L_{aq} \frac{ d \beta}{dx} \label{eq:ons}\\
j_q &=& -L_{qa} \beta\frac{d \mu}{dx} + L_{qq} \frac{ d \beta}{dx}  \quad,\nonumber
\end{eqnarray}
 where the coefficients $L_{**}$  are the entries of the symmetric and positive-definite Onsager matrix $\mathbb{L}$  and the two gradient
 terms identify the thermodynamic forces, with $\beta=1/T$ and $x=n/N$. Within this representation, coupled transport is clearly 
related to  non-vanishing off-diagonal elements $L_{aq}=L_{qa}$. Indeed, the Seebeck coefficient, that quantifies the strength
 of the coupling between the two currents, is defined as~\cite{Iubini2016}
 \begin{equation}
 S = \beta  \frac{L_{aq}}{L_{aa}} \label{eq:seeb}
 \end{equation}
and the thermodiffusive conversion efficiency is determined by the figure of merit~\cite{benenti2017fundamental}
   \begin{equation}
 ZT = \frac{L_{aq}^2}{det\,\mathbb{L}}\quad. \label{eq:ZT}
 \end{equation}

  Steady states are additionally described  in terms of the (averaged) profiles of local norm  and local energy
 \begin{eqnarray}
 a_n&=&\langle|z_n|^2\rangle\label{eq:pra} \\
 h_n&=&\langle e_n \label{eq:prh}\rangle
 \end{eqnarray}
 or, equivalently, by means of 
 local measurements of temperature $T_n$ and chemical potential $\mu_n$. For these two last quantities,
 suitable explicit microcanonical definitions 
 can be consistently derived from the thermodynamic relations $1/T = \partial s / \partial h$ and 
 $\mu/T=-\partial s / \partial a$, where $s$ is the microcanonical entropy density of the system. I refer to~\cite{Franzosi2011b,iubini2012} and references therein for details.

Numerical simulations of the dynamics of the stochastic DS equation were  performed by implementing a 4th-order Runge-Kutta
scheme with minimum timestep $\delta t =  10^{-4}$ time units. 
Without any loss of generality, the Langevin  coupling parameter $\gamma$ has  been set to 1.
In order to sample accurately the nonequilibrium steady states, the system was evolved for a time interval $t_s$ equal to $5\times10^6$
time units after a transient evolution of $10^6$ time units. It was verified that $t_s$ was long enough to observe stationary
currents and profiles in the whole range of thermodynamic parameters and for the system sizes explored in this paper.

 \section{Steady states} 
\begin{figure}[ht]
\begin{center}
\includegraphics*[width=0.8\textwidth]{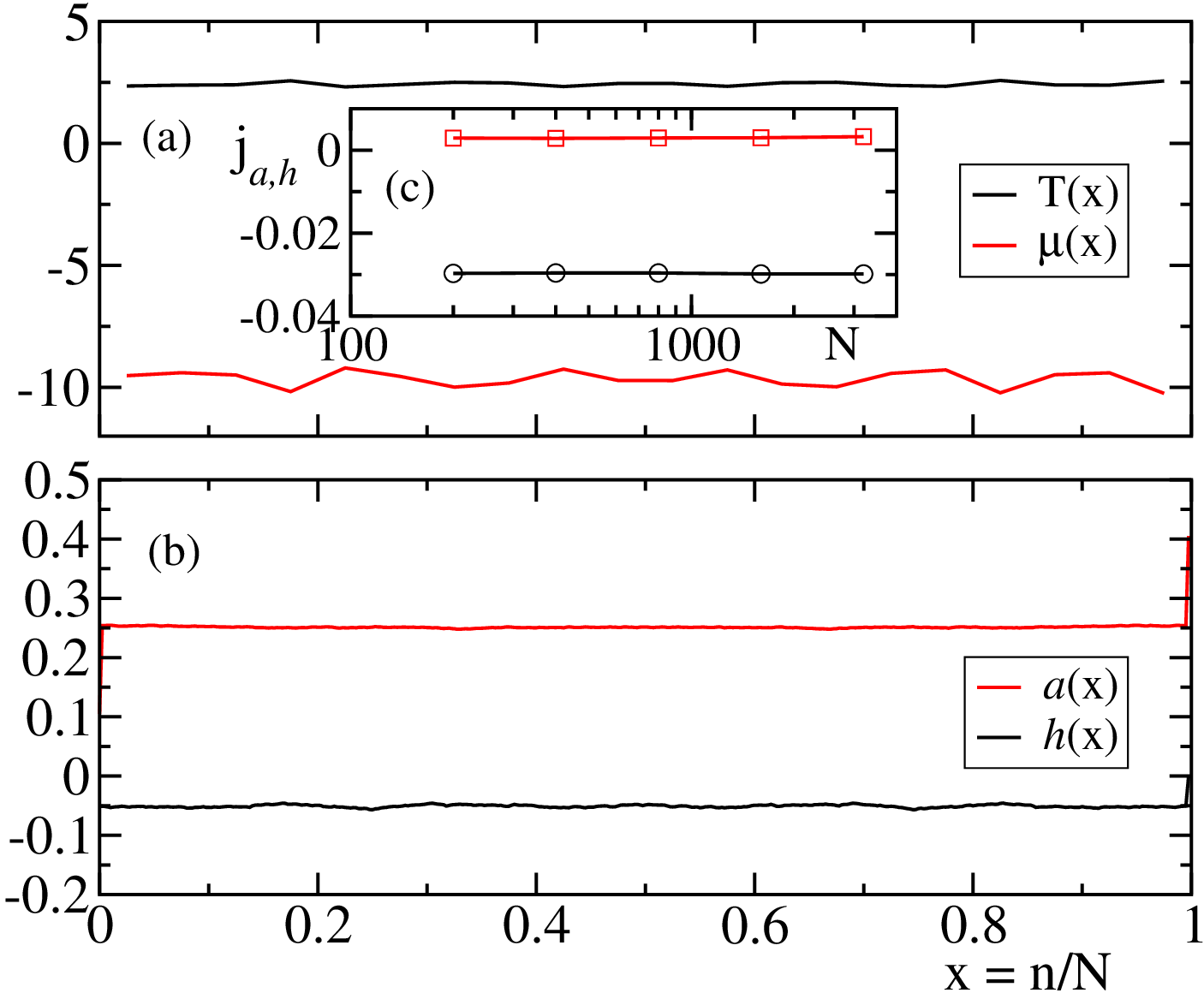}
\end{center}
\caption{Stationary nonequilibrium profiles in the integrable limit $\gamma_c=0$ as a function of $x=n/N$ for a chain of $N=400$ lattice sites and
$\mu_L=\mu_R=-10$, $T_L=1$, $T_R=4$. (a) Temperature and chemical potential profiles computed with the appropriate
microcanonical observables over small spatial subchains of 10 sites around $n$. (b) Local norm and energy densities
corresponding to the profiles in (a).  (c) Fluxes of norm (circles) and energy (squares) versus $N$ for the same thermodynamic parameters.
}
\label{f:fig2}
\end{figure}

  In this section, I discuss the transport properties of the stochastic DS chain.
   As a preliminary test for the consistency of the nonequilibrium setup, it was verified that for $\gamma_c=0$ the chain displays
  ballistic transport and flat profiles, as shown in figure~\ref{f:fig2} for a pure temperature unbalance. Similarly to the behaviour of the chain of real 
  oscillators~\cite{RLL67}, the temperature of the chain settles to a constant value which is the average between $T_L$ and $T_R$. 
  The same occurs for the chemical potential. 

\begin{figure}
\begin{center}
\includegraphics*[width=0.8\textwidth]{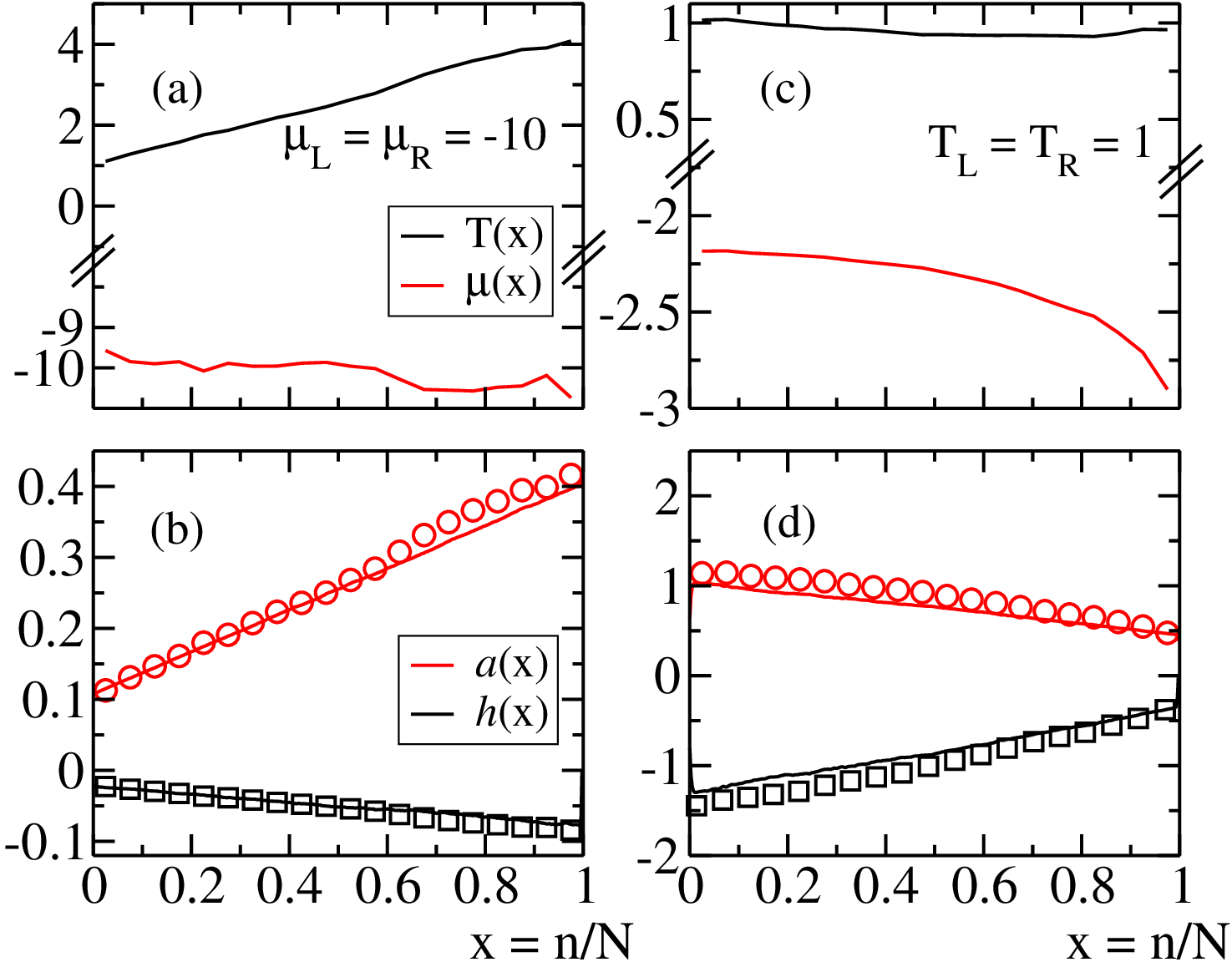}
\end{center}
\caption{
Solid lines: stationary nonequilibrium profiles
of temperature and chemical potential (upper row) and energy and norm (lower row)
 for a purely thermal unbalance [panels (a) and  (b), $T_L=1$, $T_R=4$, $\mu_L=\mu_R=-10$ ]
 and a purely chemical unbalance  [panels (c) and (d), $T_L=T_R=1$, $\mu_L=-2.2$, $\mu_R=-3$] in a chain with $N=400$. Red dots and
 black squares in panels (b) and (d) correspond to the values of local densities reconstructed from Eqs. (\ref{eq:s2}) using the local values of $T(x)$
 and $\mu(x)$ shown in panels (a) and (c), respectively. 
}
\label{f:fig3}
\end{figure}

For finite $\gamma_c$, the response of the system is radically different, as shown in figure~\ref{f:fig3} for $\gamma_c=1$.
   The linear-stochastic chain displays non-flat profiles of temperature and chemical potential 
  that interpolate between the values imposed by the external reservoirs. Figures~\ref{f:fig3}(a,b) show  the stationary profiles obtained with the same choice of boundary conditions used for the integrable limit of figure~\ref{f:fig2}.
  Depending on the choice of  the external thermal and chemical unbalances, also nonlinear profiles may emerge, as shown in 
  figure \ref{f:fig3}(c,d) for a pure chemical unbalance, see in particular the profile of $\mu(x)$.
  More in general, if the system is locally in equilibrium, a consistent representation of the stationary profiles requires that the relations  (\ref{eq:s2}) are satisfied for every $x\in[0,1]$. In figures~\ref{f:fig3}(b,d)  it is verified that this is indeed the case (up to numerical precision) by  computing $a(x)$ and $h(x)$ in two independent ways: (\textit{i}) from the direct computation through the definitions (\ref{eq:pra}) and (\ref{eq:prh}), respectively (solid lines); (\textit{ii}) from the profiles of $T(x)$ and $\mu(x)$ and using the equations of state 
  (\ref{eq:s2}) (open symbols).  These profiles are also reported parametrically in the  phase diagram $(a,h)$,
  where they give rise to almost straight paths in  the region $R_p$, see  figure~\ref{f:fig1}. 
   
\begin{figure}[ht]
\begin{center}
\includegraphics*[width=0.8\textwidth]{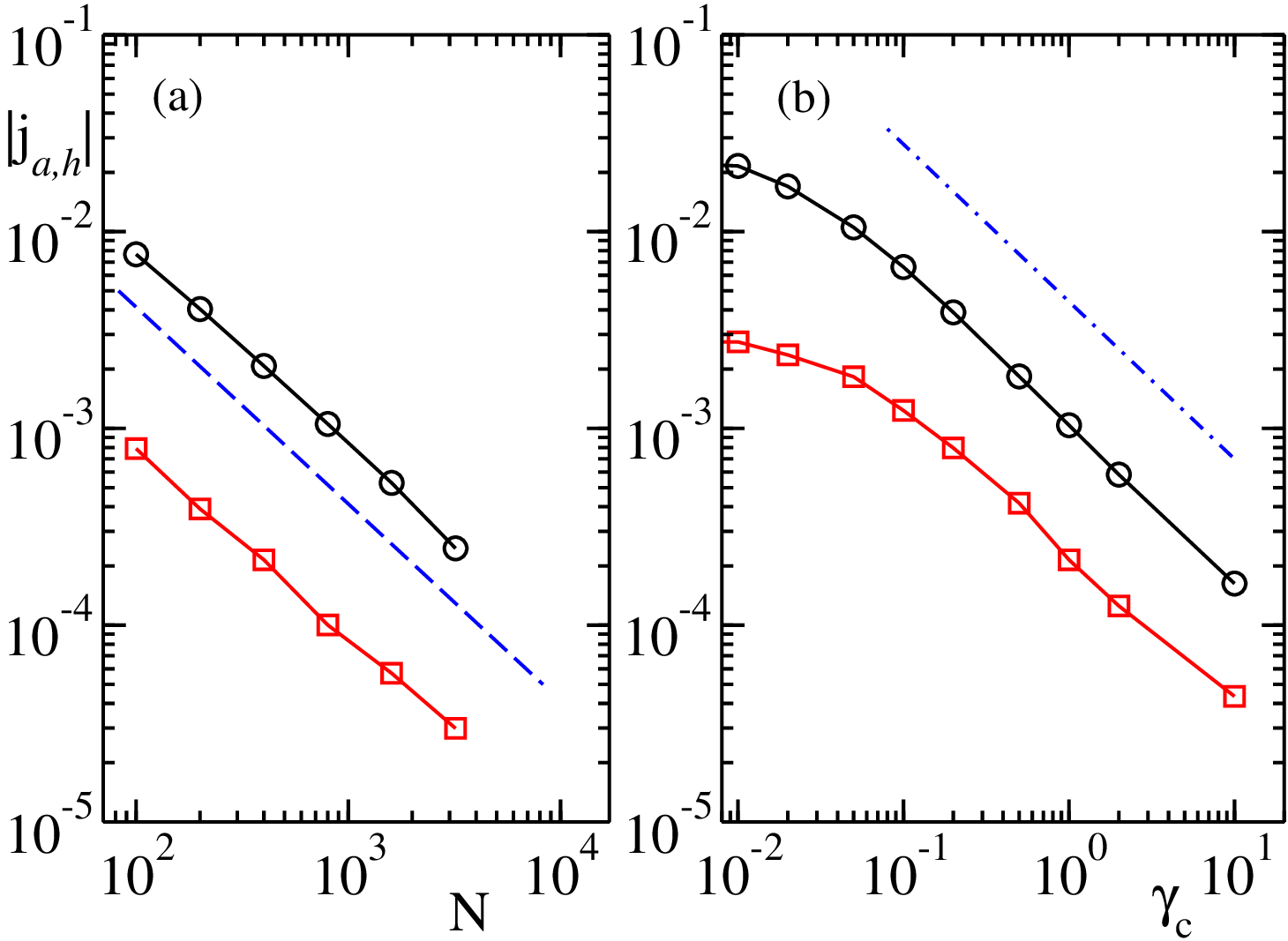}
\end{center}
\caption{
(a) Average norm current (circles) and energy current (squares) as a function of the chain length $N$ for
$T_L=1$, $T_R=4$, $\mu_L=\mu_R=-10$ and $\gamma_c=1$ 
The  dashed line shows the typical $1/N$ scaling expected from
normal transport.  (b) Same analysis for fixed $N=400$ and different values of the noise rate $\gamma_c$. 
The dot-dashed line refers to a slope $j\sim \gamma_c^{-0.8}$.
}
\label{f:fig4}
\end{figure}

For fixed differences of the thermodynamic parameters $\Delta T =T_R-T_L$ and $\Delta \mu=\mu_R-\mu_L$, the two currents $j_a$ and $j_h$ of the stochastic
DS chain  are inversely proportional to the system size $N$, as shown in figure~\ref{f:fig4}(a).
These results confirm that  transport  is normal and that the Onsager coefficients are finite in the thermodynamic limit, 
in analogy to what was observed in the presence of nonlinearities~\cite{iubini2012}.   
In any case, although the diffusive scaling holds in principle for every finite $\gamma_c$, the magnitude of stationary currents
is found to depend on $\gamma_c$, see figure~\ref{f:fig4}(b). In particular, the observed trends indicate that
in the limit of very frequent stochastic moves, transport is effectively suppressed   as a consequence of the dynamic decoupling
of Schr\"odinger oscillators. This effect is qualitatively similar to
the behaviour of quantum systems subjected to strong dephasing noise, where the suppression of unitary evolution 
is often referred to as the quantum Zeno effect~\cite{misra1977zeno}.
On a more quantitative level, the observed decay $j\sim \gamma_c^\alpha$ with $\alpha\simeq -0.8$ (dot-dashed line in figure~\ref{f:fig4}(b)) appears quite different
from the result $\alpha=-1$ expected for pure-dephasing white noise~\cite{SCHWARZER1972}. A detailed analysis of this problem   would demand  a careful study of the statistical properties of the conservative noise, a task that goes beyond the aims of the present paper.

 \begin{figure}[ht]
\begin{center}
\includegraphics*[width=0.8\textwidth]{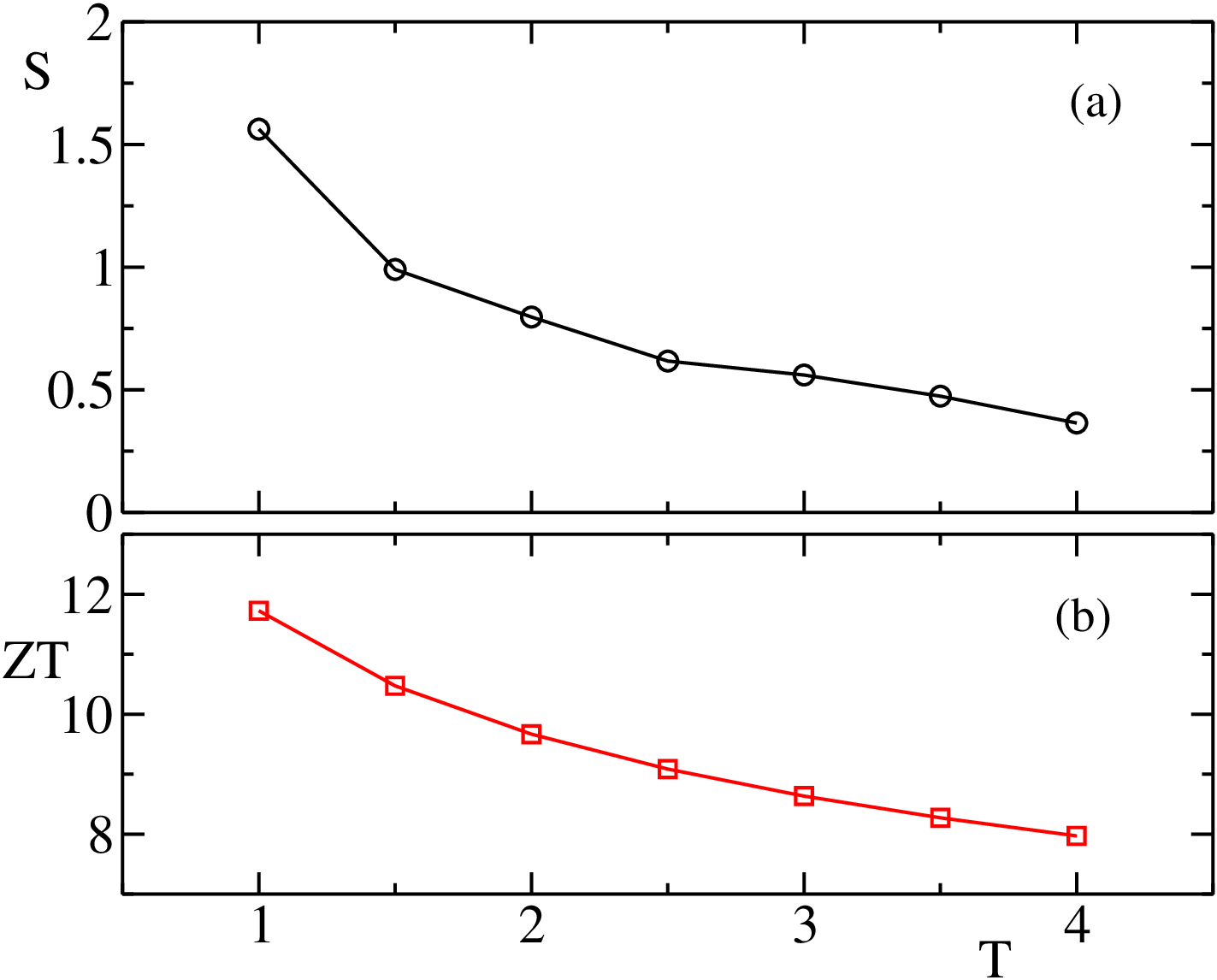}
\end{center}
\caption{
Seebeck coefficient [panel (a)] and figure of merit [panel (b)]  versus temperature $T$ along the isochemical line $\mu=-4$ for a chain of  $N=400$ lattice sites.
$S$ is derived from equation ~(\ref{eq:seeb}) and the Onsager coefficients are obtained from two series of
 nonequilibrium simulations with fixed differences $(\Delta T=0.4,\Delta \mu=0)$ and $(\Delta T =0,\Delta \mu = 0.4)$.
}
\label{f:fig5}
\end{figure}
Finally, a numerical evaluation of the Seebeck coefficient of the system was performed on the isochemical line $\mu=-4$ for temperatures in the range $1\leq T \leq 4$ (and local norms of order 1),  see figure~\ref{f:fig5}(a).  
The value of $S(\mu,T)$ was obtained from the Onsager matrix through equation~(\ref{eq:seeb}),
 see   references~\cite{iubini2012,Iubini2016} for details. In short, given a reference
 state $(T,\mu)$, the Onsager matrix was computed by applying fixed thermal and chemical unbalances to this state and  by inverting equation (\ref{eq:ons}). 
 If these unbalances are small enough, the profiles of $T(x)$ and $\mu(x)$ are  linear and  the resulting Onsager matrix is independent on the system size $N$.     
It was verified for two chains lengths $N=200$ and $N=400$ that  the choice  $\Delta T=0.4$ and $\Delta \mu=0.4$ is compatible with
 linear response in the specified region of parameters (data not shown). Accordingly, figure~\ref{f:fig5}  displays the
 results obtained with $N=400$.
 It was also checked that, within  statistical errors, the Onsager matrix is symmetric and positive-definite. 
  Altogether, the finite and positive Seebeck coefficient confirms that the stochastic DS displays coupled transport. 
  Moreover, the decreasing character of  $S(T)$ in figure~\ref{f:fig5}(a) suggests that in the limit of very high temperatures, coupled transport is suppressed 
  by the incoherent nature of the oscillators' dynamics. 
   For the same thermodynamical parameters, the figure of merit $ZT$  obtained from equation (\ref{eq:ZT})  is shown in  figure~\ref{f:fig5}(b).    It essentially follows  the behaviour of $S(T)$, with a moderate  decrease from the value $ZT\simeq 12$ attained for $T=1$.

  \section{Conclusions}
In the present work, it was introduced a linear-stochastic model of coupled complex oscillators that displays coupled transport in the sense of linear irreversible thermodynamics.
The model consists in a one-dimensional discrete Schr\"odinger chain endowed with stochastic moves that act on the local phases
and conserve simultaneously
the total energy and the total norm. Due to the linear nature of the deterministic dynamics, the equilibrium properties of the system are fully
accessible through an equation of state that was derived within the grand-canonical formalism~\cite{Rumpf2004}. 
The resulting equilibrium  phase diagram is two-dimensional and  represents the limit of low norm densities	 of the analogus diagram obtained for the DNLS  equation~\cite{Rasmussen2000}.
Nonequilibrium stationary states  crucially depend on the stochastic dynamics: despite the intrinsically discrete character of the 
conservative moves, I have shown that they are sufficient to introduce irreversibility in the system. Indeed, numerical simulations indicate that such a model displays diffusive transport and a finite Seebeck coefficient, similarly to what is found in the DNLS equation~\cite{iubini2012,Iubini2016}.



An interesting open question that naturally arises in this context concerns the possibility of performing an analytical description
  of the nonequilibrium problem 
for the stochastic Schr\"odinger  model, in analogy to what was done for the chain of real harmonic oscillators~\cite{Lepri2010}.
Practically, this task requires to find the solution of a Fokker-Planck equation, where a purely dynamical propagator is combined with
the contribution of conservative collisions~\cite{Lepri2010}. The importance of such an approach
is clearly related to the opportunity of understanding and controlling  the mechanisms of  conversion efficiency in interacting systems.

\ack
I acknowledge support from Progetto di Ricerca Dipartimentale BIRD173122/17 of the University of Padova. I thank A. Politi and S. Lepri for 
fruitful discussions.

\section*{References}

\providecommand{\newblock}{}


\begin{thebibliography}{10}
\expandafter\ifx\csname url\endcsname\relax
  \def\url#1{{\tt #1}}\fi
\expandafter\ifx\csname urlprefix\endcsname\relax\def\urlprefix{URL }\fi
\providecommand{\eprint}[2][]{\url{#2}}

\bibitem{benenti2017fundamental}
Benenti G, Casati G, Saito K and Whitney R~S 2017 {\em Physics Reports\/} {\bf
  694} 1--124

\bibitem{luo2018thermodynamic}
Luo R, Benenti G, Casati G and Wang J 2018 {\em Physical review letters\/} {\bf
  121} 080602

\bibitem{benenti2013conservation}
Benenti G, Casati G and Wang J 2013 {\em Physical review letters\/} {\bf 110}
  070604

\bibitem{mejia2001coupled}
Mejia-Monasterio C, Larralde H and Leyvraz F 2001 {\em Physical review
  letters\/} {\bf 86} 5417

\bibitem{larralde2003transport}
Larralde H, Leyvraz F and Mejia-Monasterio C 2003 {\em Journal of statistical
  physics\/} {\bf 113} 197--231

\bibitem{Lepri2016}
Lepri S (ed) 2016 {\em Thermal transport in low dimensions: from statistical
  physics to nanoscale heat transfer\/} ({\em Lect. Notes Phys\/} vol 921)
  (Springer-Verlag, Berlin Heidelberg)

\bibitem{LLP03}
Lepri S, Livi R and Politi A 2003 {\em Phys. Rep.\/} {\bf 377} 1

\bibitem{DHARREV}
Dhar A 2008 {\em Adv. Phys.\/} {\bf 57} 457--537

\bibitem{Kevrekidis}
Kevrekidis P~G 2009 {\em The Discrete Nonlinear Schr\"odinger Equation\/}
  (Springer Verlag, Berlin)

\bibitem{jensen1982}
Jensen S 1982 {\em Quantum Electronics, IEEE Journal of\/} {\bf 18} 1580--1583

\bibitem{christodoulides1988}
Christodoulides D and Joseph R 1988 {\em Optics Letters\/} {\bf 13} 794--796

\bibitem{trombettoni2001}
Trombettoni A and Smerzi A 2001 {\em Phys. Rev. Lett.\/} {\bf 86} 2353

\bibitem{livi2006}
Livi R, Franzosi R and Oppo G~L 2006 {\em Phys. Rev. Lett.\/} {\bf 97}
  060401--060401

\bibitem{hennig2010transfer}
Hennig H, Dorignac J and Campbell D~K 2010 {\em Phys. Rev. A\/} {\bf 82} 053604

\bibitem{borlenghi14}
Borlenghi S, Wang W, Fangohr H, Bergqvist L and Delin A 2014 {\em Phys. Rev.
  Lett.\/} {\bf 112} 047203

\bibitem{borlenghi15}
Borlenghi S, Iubini S, Lepri S, Chico J, Bergqvist L, Delin A and Fransson J
  2015 {\em Phys. Rev. E\/} {\bf 92} 012116

\bibitem{iubini2012}
Iubini S, Lepri S and Politi A 2012 {\em Physical Review E\/} {\bf 86} 011108

\bibitem{iubini2013}
Iubini S, Franzosi R, Livi R, Oppo G~L and Politi A 2013 {\em New Journal of
  Physics\/} {\bf 15} 023032

\bibitem{Iubini2016}
Iubini S, Lepri S, Livi R and Politi A 2016 {\em New J. Phys.\/} {\bf 18}
  083023

\bibitem{Mendl2015}
Mendl C~B and Spohn H 2015 {\em J. Stat. Mech: Theory Exp.\/} {\bf 2015} P08028

\bibitem{iubini2014}
Iubini S, Lepri S, Livi R and Politi A 2014 {\em Physical review letters\/}
  {\bf 112} 134101

\bibitem{wall17}
Iubini S, Lepri S, Livi R, Oppo G~L and Politi A 2017 {\em Entropy\/} {\bf 19}
  445

\bibitem{RLL67}
Rieder Z, Lebowitz J~L and Lieb E 1967 {\em J. Math. Phys.\/} {\bf 8} 1073

\bibitem{sheng2006}
Sheng P 2006 {\em {Introduction to wave scattering, localization, and
  mesoscopic phenomena}\/} vol~88 (Springer)

\bibitem{Dhar2006}
Dhar A and Roy D 2006 {\em Journal of Statistical Physics\/} {\bf 125} 801--820

\bibitem{BBO06}
Basile G, Bernardin C and Olla S 2006 {\em Phys. Rev. Lett.\/} {\bf 96} 204303

\bibitem{DLLP08}
Delfini L, Lepri S, Livi R and Politi A 2008 {\em Phys. Rev. Lett.\/} {\bf 101}
  120604

\bibitem{Lepri2009}
Lepri S, Mej{\'\i}a-Monasterio C and Politi A 2009 {\em J. Phys. A: Math.
  Theor.\/} {\bf 42} 025001

\bibitem{Lepri2010}
Lepri S, Mej{\'\i}a-Monasterio C and Politi A 2010 {\em Journal of Physics A:
  Mathematical and Theoretical\/} {\bf 43} 065002

\bibitem{Delfini10}
Delfini L, Lepri S, Livi R, Mejía-Monasterio C and Politi A 2010 {\em J. Phys.
  A: Math. Theor.\/} {\bf 43} 145001

\bibitem{spohn2014fluctuating}
Spohn H 2014 {\em arXiv preprint arXiv:1411.3907\/}

\bibitem{olla2019fourier}
Olla S 2019 {\em arXiv preprint arXiv:1905.07762\/}

\bibitem{letizia2017diffusive}
Letizia V 2017 {\em arXiv preprint arXiv:1712.03590\/}

\bibitem{Rumpf2004}
Rumpf B 2004 {\em Phys. Rev. E\/} {\bf 69} 016618

\bibitem{Rasmussen2000}
Rasmussen K, Cretegny T, Kevrekidis P and Gr{\o}nbech-Jensen N 2000 {\em Phys.
  Rev. Lett.\/} {\bf 84} 3740--3743

\bibitem{ramsey1956}
Ramsey N~F 1956 {\em Physical Review\/} {\bf 103} 20

\bibitem{Saito2010}
Saito K, Benenti G and Casati G 2010 {\em Chem. Phys.\/} {\bf 375} 508--513

\bibitem{Franzosi2011b}
Franzosi R 2011 {\em J. Stat. Phys.\/} {\bf 143}(4) 824--830

\bibitem{misra1977zeno}
Misra B and Sudarshan E~G 1977 {\em Journal of Mathematical Physics\/} {\bf 18}
  756--763

\bibitem{SCHWARZER1972}
Schwarzer E and Haken H 1972 {\em Physics Letters A\/} {\bf 42} 317 -- 318

\end{thebibliography}
\end{document}